\documentclass[aps,pre,preprint,showpacs]{revtex4}
\usepackage{bm}
\usepackage{graphicx}
\begin{document}
\title{Steady state representation of the homogeneous cooling
state of a granular gas}
\author{J. Javier Brey}
\author{M.J. Ruiz--Montero}
\author{F. Moreno}
\affiliation{F\'{\i}sica Te\'{o}rica, Universidad de Sevilla,
Apdo.\ de
 Correos 1065, E-41080 Sevilla, Spain}
\date{today}

\begin{abstract}
The properties of a dilute granular gas in the homogeneous cooling
state are mapped to those of a stationary state by means of a
change in the time scale that does not involve any internal
property of the system. The new representation is closely related
with a general property of the granular temperature in the long
time limit. The physical and practical implications of the mapping
are discussed. In particular, simulation results obtained by the
direct simulation Monte Carlo method applied to the scaled
dynamics are reported. This includes ensemble averages and also
the velocity autocorrelation function, as well as the
self-diffusion coefficient obtained from the latter by means of
the Green-Kubo representation. In all cases, the obtained results
are compared with theoretical predictions.

\end{abstract}

\maketitle

\section{Introduction}
\label{s1} A granular fluid is a collection of macroscopic
particles interacting via short range hard inelastic collisions.
Particles move in a ballistic way between collisions and total
momentum is conserved \cite{PyL01}. The prototypical idealized
model for granular fluids is a system of inelastic smooth hard
spheres or disks, with the inelasticity of collisions being
described by means of a constant (independent of the relative
velocity) coefficient of normal restitution. Then, in the last
years the traditional methods of kinetic theory and
non-equilibrium statistical mechanics have been extended to the
case of inelastic collisions. Quite remarkably, it has been
realized that the single feature of incorporating energy
dissipation in collisions is able to provide a theoretical scheme
where many of the peculiar features exhibited by real granular
fluids can be tackled. This includes phenomena such as the
development of strong density and temperature inhomogeneities that
are not induced by the boundary conditions \cite{GyZ93,KWyG97},
spontaneous symmetry breaking in partitioned \cite{SyN96,BMGyR02}
and non-partitioned systems \cite{LMyS02}, segregation in systems
composed of different kind of particles \cite{RVyP91}, and pattern
formation \cite{UMyS96}, to cite a few examples. In most of these
cases, the usefulness of a collective description of the system in
terms of hydrodynamic-like equations has been verified. Such a
description can only be fully understood and justified by starting
from a more fundamental particle level, as considered in kinetic
theory.

Due to energy dissipation in collisions, granular systems do not
present a stationary, homogeneous and isotropic state similar to
the equilibrium one of ordinary fluids. The simplest possible
state corresponds to a freely evolving homogenous and isotropic
system whose energy decays monotonically in time, the so-called
homogeneous cooling state (HCS). This state plays a relevant role
in order to investigate the transport properties of a granular
fluid, since it provides the zeroth order in the gradients
approximation when applying the Chapman-Enskog procedure to derive
hydrodynamics from a given kinetic equation \cite{BDKyS98,GyD99}.
Also, linear response around this state has been studied and
formal expressions for the Navier-Stokes transport coefficients
have been derived \cite{GyvN00,DyB02}. They are the generalization
to inelastic systems of the well-known Green-Kubo formulas.
Besides their theoretical interest, they allow a direct
determination of the transport coefficients from the dynamics of
the system in the HCS, without introducing any additional
approximation, by using numerical simulation methods.

Molecular dynamics (MD) simulation provides a method to
investigate a system of particles  at the most fundamental level
of description. Nevertheless, when applied to a granular fluid in
the HCS, several limitations show up. First, since the system is
continuously cooling, the typical velocity of the particles
becomes very small rather soon and numerical inaccuracies become
very large. In principle, this could be solved by introducing some
kind of external thermostat, but then the relationship between the
original HCS and the state being actually simulated is not clear.
Another possibility is to take advantage of  the fact that there
is no intrinsic time scale in a system of hard particles, and to
rescale the velocity of all particles after every collision, so
that the energy is forced to remain constant \cite{SyM01}.
Although it seems that this method must lead to correct results
for time-independent properties of the HCS, e.g. structural
properties or the own scaled velocity distribution, in the
infinite system limit, it is not evident how to extract from the
simulation data properties of the actual dynamics of the system
involving time fluctuations or two-time correlations.

Very recently, a procedure has been introduced according to which
the dynamics of the system in the HCS is exactly mapped onto the
dynamics around a steady state by means of a change in the time
scale being used \cite{Lu01}. The change is independent of the
state of the system. This is possible because the temperature of
the HCS becomes independent of its initial value in the long time
limit. This is a very strong and fundamental property of that
state that has not received too much attention up to now. Then,
the existence of the steady state in the scaled time
representation is tied to the own physical properties of the
mechanism of energy dissipation.

The second limitation of the MD simulation of a granular fluid in
the HCS is associated with the fact that this state is unstable
with respect to spacial long wavelength perturbations
\cite{GyZ93,McyY94}. This instability has been identified in the
context of hydrodynamics, and an expression for the critical size
of the system, beyond which it becomes unstable, has been
obtained. The critical size is a function of the density and the
coefficient of restitution, decreasing with the former and
increasing with the latter. In practice, this implies that for
high densities and/or small values of the coefficient of
restitution, only very small systems can be simulated in the HCS,
and undesired finite size effects might influence the results.

The aim of this paper is to investigate in detail the physical and
practical implications of the steady state representation of the
HCS mentioned above for a low density granular gas. Attention will
be paid not only to the one-time properties of the system, but
also to two-time correlation functions. While the former can be
discussed on the basis of the inelastic Boltzmann equation, the
analysis of the correlation functions requires to introduce an
equation for the two-particle and two-time distribution function.
This is done by a direct extension of the methods used in the
elastic case for out of equilibrium systems \cite{EyC81}. It is
shown that both kind of properties can be expressed in terms of
averages over the stationary state of the system. Special emphasis
will be put on the relationship between the theoretical
description of the system in terms of reduced distribution
functions in the low density limit and the underlying $N$-particle
dynamics. This is important in order to implement the calculation
of a given property by means of the direct simulation Monte Carlo
method (DSMC) \cite{Bi94}. It must be kept in mind that this
method is designed not just as a numerical tool to solve the
Boltzmann equation, but as a real $N$-particle dynamics simulation
of a low density gas. In this sense, it is expected to provide not
only the one-particle distribution function of the system, but its
complete dynamical description.

One of the main practical advantages of the DSMC method is that it
allows to incorporate at the level of the particle dynamics the
spatial symmetry properties of the particular physical situation
of interest. For instance, it is easy to restrict the dynamics of
the system so that it remains homogeneous, with no possibility of
developing spacial inhomogeneities. In this way, the second
limitation of MD simulations of the HCS can be overcome for a low
density gas with no restriction on the size of the system or the
number of particles being used. The combination of this feature of
the DSMC method and the steady representation of the HCS offers an
almost unique way to investigate the properties of a dilute
granular gas.

The plan of the paper is as follows. In Sec. \ref{s2}, the
Boltzmann equation and some of the results for the one-particle
distribution function of the HCS are shortly reviewed. The
peculiar long time behavior of the temperature is indicated, and
in Sec.\ \ref{s3} it is used to construct a representation of the
dynamics of the system in which the distribution function of the
HCS becomes time independent. It is shown that the asymptotic
steady temperature is determined by an intrinsic property of the
system, namely the cooling rate. Then, dynamical simulations of a
system of inelastic hard disks with the DSMC method are presented,
and the numerical results for the cooling rate obtained from the
values of the steady temperature are compared with the existing
theoretical predictions. Results for the fourth moment of the
velocity distribution function are also reported. They are
consistent with those obtained previously by using the original
dynamics, in which the HCS is time dependent.

Time correlation functions of dynamical variables in the HCS are
addressed in Sec.\ \ref{s4}, where their low density limit is
analyzed and the relationship between their values in the original
and scaled dynamics is established. The particular case of the
correlation functions appearing in the Green-Kubo form of the
transport coefficients of a low density gas \cite{DyB02} is
considered. As an example, the velocity autocorrelation function,
and from it the self-diffusion coefficient, are computed in Sec.
\ref{s5} for a system of inelastic hard disks, and the results are
compared with the theoretical predictions obtained by the
Chapman-Enskog method in the first Sonine approximation. The paper
ends with a summary and a brief discussion.

\section{Basic equations and the homogeneous cooling state}
\label{s2} We consider a system of $N$ inelastic hard spheres
($d=3$) or disks ($d=2$) of mass $m$ and diameter $\sigma$. The
position and velocity coordinates of particle $i$ at time $t$ will
be represented by ${\bm R}_{i}(t)$ and ${\bm V}_{i}(t)$,
respectively. The particle dynamics consists of free streaming
until a pair of particles $i$ and $j$ are at contact and a
collision takes place. The effect of the collision is to
instantaneously change the velocities of the two involved
particles according to
\begin{eqnarray}
\label{2.1} {\bm V}_{i} \rightarrow {\bm V}^{\prime}_{i} \equiv
b_{\bm \sigma} {\bm V}_{i} & = & {\bm V}_{i} -\frac{1+\alpha}{2}
\left( \widehat{\bm \sigma} \cdot {\bm V}_{ij} \right)
\widehat{\bm \sigma} \nonumber \\
{\bm V}_{j} \rightarrow {\bm V}^{\prime}_{j} \equiv b_{\bm \sigma}
{\bm V}_{j} & = & {\bm V}_{j} +\frac{1+\alpha}{2} \left(
\widehat{\bm \sigma} \cdot {\bm V}_{ij} \right) \widehat{\bm
\sigma},
\end{eqnarray}
where ${\bm V}_{ij}={\bm V}_{i}-{\bm V}_{j}$ is the relative
velocity, $\widehat{{\bm \sigma}}$ is the unit vector pointing
from the center of particle $j$ to the center of particle $i$ at
contact, and $\alpha$ is the coefficient of normal restitution. It
is defined in the range $0 < \alpha \leq 1$ and will be considered
as velocity independent along this work.

In the low density limit, it is assumed that the time evolution of
the one-particle distribution function of the system $f({\bm
r},{\bm v},t)$ is accurately described by the inelastic Boltzmann
equation \cite{GyS95,BDyS97}
\begin{equation}
\label{2.2} \left( \frac{\partial}{\partial t} + {\bm v} \cdot
\frac{\partial}{\partial {\bm r}} \right) f({\bm r},{\bm v},t) =
J[f,f],
\end{equation}
with the inelastic Boltzmann collision operator given by
\begin{equation}
\label{2.3} J[f,f] \equiv \sigma^{d-1} \int d{\bm v}_{1} \int d
\widehat{\bm \sigma}\,  \Theta ({\bm g} \cdot \widehat{\bm
\sigma}) {\bm g} \cdot \widehat{\bm \sigma} \left( \alpha^{-2}
b_{\bm \sigma}^{-1} -1 \right) f({\bm r},{\bm v},t) f({\bm r},{\bm
v}_{1},t).
\end{equation}
Here ${\bm g}={\bm v}-{\bm v}_{1}$, $\Theta$ is the Heaviside step
function, and $b_{\bm \sigma}^{-1}({\bm v},{\bm v}_{1})$ is an
operator replacing all the velocities ${\bm v}$ and ${\bm v}_{1}$
appearing to its right by the precollisional values ${\bm v}^{*}$
and ${\bm v}_{1}^{*}$ given by
\begin{eqnarray}
\label{2.4} {\bm v}^{*} & \equiv & b_{\bm \sigma}^{-1} {\bm v} =
{\bm v} -\frac{1+\alpha}{2 \alpha} (\widehat{{\bm \sigma}} \cdot
{\bm
g}) \widehat{\bm \sigma},  \nonumber \\
{\bm v}^{*}_{1} & \equiv & b_{\bm \sigma}^{-1} {\bm v}_{1} =  {\bm
v}_{1} + \frac{1+\alpha}{2 \alpha} (\widehat{{\bm \sigma}} \cdot
{\bm g}) \widehat{\bm \sigma}.
\end{eqnarray}
We are using lower-case symbols to represent field variables as
those appearing in the one-particle distribution function, while
capital symbols are used for the particle variables. Equation
(\ref{2.2}) has a particular solution describing the HCS and
having the scaling property \cite{GyS95}
\begin{equation}
\label{2.5} f_{HCS}({\bm v},t)=n_{H} v_{0}^{-d}(t) \chi_{HCS}
({\bm c}),
\end{equation}
where
\begin{equation}
\label{2.6} v_{0}(t)= \left[\frac{2k_{B}T_{HCS}(t)}{m}
\right]^{1/2}
\end{equation}
is the thermal velocity and $\chi_{HCS}({\bm c})$ is an isotropic
function of
\begin{equation}
\label{2.7} {\bm c}=\frac{\bm v}{v_{0}(t)}\, .
\end{equation}
In Eq. (\ref{2.6}), $k_{B}$ is the Boltzmann constant that is
usually set equal to unity in the literature of granular flows. We
prefer to keep it here for dimensional reasons. The number of
particles density $n$ and the granular temperature $T(t)$ are
defined, for arbitrary situations, in terms of the one-particle
distribution function in the usual way,
\begin{equation}
\label{2.8} n({\bm r},{\bm t})= \int d{\bm v} f({\bm r},{\bm
v},t),
\end{equation}
\begin{equation}
\label{2.9} \frac{d}{2}n({\bm r},t) k_{B} T({\bm r},t)=\int d{\bm
v} \frac{1}{2} m ({\bm v}-{\bm u})^{2} f({\bm r},{\bm v},t),
\end{equation}
\begin{equation}
\label{2.9a} n({\bm r},t) {\bm u}({\bm r},t) = \int d{\bm v} {\bm
v} f({\bm r}, {\bm v},t).
\end{equation}

Therefore, in the HCS all the time dependence of the distribution
function occurs through the temperature $T_{HCS}(t)$, whose
evolution equation is easily obtained from the own Boltzmann
equation
\begin{equation}
\label{2.10} \frac{\partial}{\partial
t}T_{HCS}(t)+\zeta_{HCS}(t)T_{HCS}(t)=0,
\end{equation}
where the cooling rate $\zeta_{HCS}$ is given by
\begin{equation}
\label{2.11}
\zeta_{HCS}(t)=\frac{(1-\alpha^{2})\pi^{\frac{d-1}{2}}
\sigma^{d-1}n_{H}v_{0}(t)}{2\,  \Gamma \left( \frac{d+3}{2}
\right)d} \int d{\bm c} \int d{\bm c}_{1}\, |{\bm c}-{\bm
c}_{1}|^{3}\chi_{HCS}({\bm c}) \chi_{HCS}({\bm c}_{1}).
\end{equation}
Then, $\zeta_{HCS}$ is proportional to $T_{HCS}^{1/2}$ and,
therefore, Eq. (\ref{2.10}) can be formally integrated to get the
explicit time dependence of the temperature in the HCS,
\begin{equation}
\label{2.12} T_{HCS}(t)=T_{HCS}(0) \left[ 1+\frac{1}{2}
\zeta_{HCS}(0) t \right]^{-2}.
\end{equation}
This expression is known as Haff's law \cite{Ha83} and it has the
interesting property of becoming independent of the initial
temperature in the long time limit. More precisely, it reduces to
\begin{equation}
\label{2.13} T_{HCS} \sim 4(\bar{\zeta}t)^{-2},
\end{equation}
with
\begin{equation}
\label{2.14}
\bar{\zeta}=\frac{\zeta_{HCS}(t)}{T_{HCS}^{1/2}(t)}=\frac{v_{0}(t)
\zeta_{0}}{\ell T_{HCS}^{1/2}(t)}\, .
\end{equation}
In the last transformation we have introduced the time-independent
dimensionless cooling rate
\begin{equation}
\label{2.15} \zeta_{0}=\frac{\ell \zeta_{HCS}(t)}{v_{0}(t)},
\end{equation}
where $\ell \equiv (n_{H}\sigma^{d-1})^{-1}$ is proportional to
the mean free path. The above long time behavior of the
temperature implies the existence of an asymptotic regime in which
the HCS becomes independent of the initial condition or, in other
words, all the homogeneous cooling states of a given system tend
to converge in the long time limit. This is an exact property
following from the own existence of the HCS and it will be
exploited in the next Section in order to introduce a steady
representation of the HCS.

Substitution of Eq.\ (\ref{2.5}) into the Boltzmann equation
(\ref{2.3}) provides a closed integro-differential equation for
the function $\chi_{HCS}({\bm c})$,
\begin{equation}
\label{2.16} \frac{\zeta_{0}}{2} \frac{\partial}{\partial {\bm c}}
\cdot \left( {\bm c}\chi_{HCS}\right)=
J_{c}[\chi_{HCS},\chi_{HCS}],
\end{equation}
\begin{equation}
\label{2.17} J_{c}[\chi_{HCS},\chi_{HCS}] = \int d{\bm c}_{1} \int
d \widehat{\bm \sigma}\, \Theta [({\bm c}-{\bm c}_{1}) \cdot
\widehat{\bm \sigma}] ({\bm c}-{\bm c}_{1}) \cdot \widehat{\bm
\sigma} \left[ \alpha^{-2} b_{\bm \sigma}^{-1}({\bm c},{\bm
c}_{1}) -1 \right] \chi_{HCS}({\bm c}) \chi_{HCS}({\bm c}_{1}).
\end{equation}
The operator $b_{\bm \sigma}^{-1}({\bm c},{\bm c}_{1})$ is again
defined by Eqs. (\ref{2.4}) but substituting the velocities ${\bm
v}$, ${\bm v}_{1}$ by ${\bm c}$, ${\bm c}_{1}$.

The solution of Eq. (\ref{2.16}) is only partially known. In
particular, an expansion in Sonine polynomials has been
considered. To first order, $\chi_{HCS}({\bm c})$ is approximated
by
\begin{equation}
\label{2.18} \chi_{HCS}({\bm c})= \frac{e^{-c^{2}}}{\pi^{d/2}}
\left[ 1+a_{2}(\alpha) S^{(2)}(c^{2}) \right],
\end{equation}
where
\begin{equation}
\label{2.19} S^{(2)}(c^{2})=\frac{c^{4}}{2}-\frac{d+2}{2}\,
c^{2}+\frac{d(d+2)}{8}.
\end{equation}
The coefficient $a_{2}(\alpha)$ turns out to be proportional to
the fourth cumulant of the distribution, namely
\begin{equation}
\label{2.20} a_{2}(\alpha)=\frac{4}{d(d+2)} \left[ \langle c^{4}
\rangle -\frac{d(d+2)}{4} \right],
\end{equation}
\begin{equation}
\label{2.21} \langle c^{4} \rangle \equiv \int d{\bm c}\, c^{4}
\chi_{HCS}({\bm c}).
\end{equation}
When Eq.\, (\ref{2.18}) is substituted into Eq.\ (\ref{2.16}) a
closed equation for $a_{2}$ is obtained. If nonlinear terms in
$a_{2}$ are neglected in this equation, it is found
\cite{GyS95,vNyE98}:
\begin{equation}
\label{2.22}
a_{2}(\alpha)=\frac{16(1-\alpha)(1-2\alpha^{2})}{9+24d+(8d-41)+30
\alpha^{2}-30 \alpha^{3}}\, .
\end{equation}
In the same approximation, Eqs.\ (\ref{2.11}) and (\ref{2.15})
yield
\begin{equation}
\label{2.23} \zeta_{0}=\frac{\sqrt{2} \pi^{\frac{d-1}{2}}
(1-\alpha^{2})}{\Gamma \left( \frac{d}{2} \right) d} \left[
1+\frac{3}{16} a_{2}(\alpha) \right].
\end{equation}

The above expression for $\chi_{HCS}({\bf c})$ is expected to be
accurate in the thermal velocity region, i.e. for velocities $c$
of the order of a few units. This has been confirmed by DSMC
simulations of a granular gas \cite{BRyC96}.

\section{Steady state representation of the HCS}
\label{s3} The long time behavior of the HCS discussed in the
preceding Section, suggests that a steady representation of it can
be obtained on a time scale $\tau$ defined by \cite{Lu01}
\begin{equation}
\label{3.1} \omega_{0} \tau =\ln \frac{t}{t_{0}}\, ,
\end{equation}
where $\omega_{0}$ and $t_{0}$ are arbitrary constants.
Consistently, the velocity ${\bm W}_{i}$ of particle $i$ is given
in the new scale by
\begin{equation}
\label{3.2} {\bm W}_{i}(\tau)=\omega_{0}t_{0}e^{\omega_{0} \tau}
{\bm V}_{i}(t) = \omega_{0} t {\bm V}_{i}(t).
\end{equation}
The particle dynamics in these variables consists of an
accelerating streaming between collisions,
\begin{eqnarray}
\label{3.3} \frac{\partial }{\partial \tau}{\bm R}_{i}(\tau)& = &
{\bm W}_{i}(\tau), \nonumber \\
\frac{\partial}{\partial \tau} {\bm W}_{i}(\tau)& = & \omega_{0}
{\bm W}_{i}(\tau),
\end{eqnarray}
while the effect of a collision between particles $i$ and $j$ is
to instantaneously modify their velocities accordingly with the
same rules as given in Eqs. (\ref{2.1}), i.e.
\begin{eqnarray}
\label{3.4} {\bm W}_{i} \rightarrow {\bm W}^{\prime}_{i} \equiv
b_{\bm \sigma} {\bm W}_{i} & = & {\bm W}_{i} -\frac{1+\alpha}{2}
\left( \widehat{\bm \sigma} \cdot {\bm W}_{ij} \right)
\widehat{\bm \sigma} \nonumber \\
{\bm W}_{j} \rightarrow {\bm W}^{\prime}_{j} \equiv b_{\bm \sigma}
{\bm W}_{j} & = & {\bm W}_{j} +\frac{1+\alpha}{2} \left(
\widehat{\bm \sigma} \cdot {\bm W}_{ij} \right) \widehat{\bm
\sigma}.
\end{eqnarray}
Of course, this invariance of the collision rules is a consequence
of the instantaneous character of the hard collisions or, in other
words, of the absence of an intrinsic time scale for hard
particles. In the $\tau$ time scale, the Boltzmann equation
(\ref{2.2}) takes the form
\begin{equation}
\label{3.5} \left( \frac{\partial}{\partial \tau}+ \omega_{0}
\frac{\partial}{\partial {\bm w}} \cdot {\bm w}+ {\bm w} \cdot
\frac{\partial}{\partial {\bm r}} \right) \tilde{f} ({\bm r},{\bm
w},t) = J[\tilde{f},\tilde{f}].
\end{equation}
Here $J$ is the same collision operator as in Eq. (\ref{2.3}) with
but substituting the velocities ${\bm v}$, ${\bm v}_{1}$ by ${\bm
w}$, ${\bm w}_{1}$, and the scaled one-particle distribution
function is
\begin{equation}
\label{3.6} \tilde{f}({\bm r},{\bm w},t) = \left( \omega_{0}t
\right)^{-d} f({\bm r},{\bm v},t).
\end{equation}

Therefore, the only modification of the Boltzmann equation is that
a new term appears in the streaming part of the equation, as
expected because of Eqs. (\ref{3.3}). This term has the same form
as some thermostats introduced more or less artificially in order
to allow the system to have a stationary state; however here it
arises solely from a change in the time scale.

Let us assume now that the system is initially in the HCS and that
it remains in it along its time evolution. This implies to
suppress the possibility that the system spontaneously develops
spatial inhomogeneities due to the existence of a long wavelength
hydrodynamic instability, the so-called clustering instability
\cite{GyZ93,McyY94}. It is worth to stress that this effect is
also present in the low density description provided by the
Boltzmann equation \cite{BRyC96,BRyC99}. In principle, the
instability can be avoided by considering small enough systems,
but this might lead to the presence of undesired finite size
effects, especially for small values of the coefficient of
restitution $\alpha$. Nevertheless, at the level of description
provided by the Boltzmann equation, this can be formally
accomplished by restricting ourselves to consider the homogeneous
form of Eq. (\ref{3.5}). Then, a scaled temperature
$\widetilde{T}_{HCS}(\tau)$ can be defined by
\begin{equation}
\label{3.7} \frac{d}{2} n_{H} k_{B} \widetilde{T}_{HCS}(\tau)=
\int d{\bm w}\, \frac{1}{2} m w^{2} \tilde{f}_{HCS}({\bm w},\tau),
\end{equation}
where $\tilde{f}_{HCS}({\bm w},\tau)$ is the scaled one-particle
distribution of the HCS. An evolution equation for this
temperature is easily obtained by using Eq.\ (\ref{2.10}),
\begin{equation}
\label{3.8} \left( \frac{\partial}{\partial \tau} - 2 \omega_{0}
\right) \widetilde{T}_{HCS}(\tau)=-\bar{\zeta}
\widetilde{T}_{HCS}^{3/2} (\tau).
\end{equation}
The solution of this equation is
\begin{equation}
\label{3.9} \widetilde{T}_{HCS}(\tau)=\left( \frac{2
\omega_{0}}{\bar{\zeta}} \right)^{2} \left[ 1+\left( \frac{2
\omega_{0}}{\bar{\zeta} \widetilde{T}_{HCS}^{1/2}(0)}-1 \right)
e^{-\omega_{0} \tau} \right]^{-2}.
\end{equation}
It follows that any initial value of the temperature tends to a
final steady value given by
\begin{equation}
\label{3.10} \widetilde{T}_{st}= \left( \frac{2
\omega_{0}}{\bar{\zeta}} \right)^{2}.
\end{equation}
As discussed above, this result only holds as long as the system
stays indefinitely in the HCS. By means of Eq.\, (\ref{2.14}), it
can also be expressed in the equivalent form
\begin{equation}
\label{3.11} \zeta_{0}=\frac{2 \omega_{0}
\ell}{\widetilde{v}_{0,st}},
\end{equation}
with
\begin{equation}
\label{3.12} \widetilde{v}_{0,st}=\left( \frac{2 k_{B}
\widetilde{T}_{st}}{m} \right)^{1/2}.
\end{equation}
Therefore, $\omega_{0}/\widetilde{v}_{0,st}$ is independent of
$\omega_{0}$ and proportional to an intrinsic property of the
system, namely the dimensionless cooling rate $\zeta_{0}$.

Moreover, once the scaled temperature has reached its steady
value, the scaled distribution function has the time independent
form
\begin{equation}
\label{3.13} \tilde{f}_{st}({\bm
w})=n_{H}\widetilde{v}_{0,st}^{-d} \chi_{HCS}(\widetilde{\bm c}),
\quad {\widetilde{\bm c}}=\frac{\bm c}{\widetilde{w}_{0,st}}=
\frac{\widetilde{v}_{0,st}}{v_{0}(t)}\, {\bm v}.
\end{equation}
Upon writing the last equality for $\widetilde{\bm c}$, we have
taken into account the asymptotic form of $T_{HCS}(t)$ given in
Eq.\ (\ref{2.13}). The existence of a steady solution of Eq.\
(\ref{3.5}) is enabled by the term proportional to $\omega_{0}$,
so that the acceleration between collisions is able to balance the
loss of energy in them. In order to avoid misunderstandings, it
must be noted that the fact that the system is in the HCS does not
impliy by itself that the temperature in the scaled variables
takes its steady value. This only happens in the long time limit.
However, what is relevant is that there is a mapping between the
steady state in the scaled representation and the associated HCS,
for arbitrary value of the parameter $\omega_{0}$. Of course, for
any arbitrary state of the system, it is possible to relate every
property in the original variables with the corresponding (scaled)
property in the scaled representation, but attention will be
restricted in the following to the HCS.

In order to confirm the practical usefulness of the steady
representation, we have carried out DSMC simulations \cite{Bi94}
of a low density granular gas whose dynamics is defined by Eqs.
(\ref{3.3}) and (\ref{3.4}). This $N$-particle simulation
algorithm is known to be consistent with the Boltzmann equation,
in the sense that it provides numerical solutions of the equation.
But the results coming from this kind of simulations go much
further and cover all the properties of a dilute gas. This
includes, in particular, fluctuations and non-equilibrium
correlations, although the precise relationship between the
simulation algorithm and the theory following from a description
of the system in the context of kinetic theory, or non-equilibrium
statistical mechanics, has not been established yet. In fact, the
method tries to mimic, by means of a Markov process, the dynamics
of a low density gas by uncoupling the streaming motion of the
particles, given by Eqs. (\ref{3.3}), and collisions during a
small enough time interval, and also by neglecting velocity
correlations in collisions. This is done independently of the
number $N$ of particles being simulated (and, therefore, of their
number density). In practice, this is implemented by dividing the
coordinate space into cells of size smaller than the mean free
path, and considering that all pairs of particles in the same cell
can collide with a probability proportional to their relative
velocity.

One of the main technical advantages of DSMC, as compared with
other particle simulation methods, is that it allows to
incorporate in the simulation algorithm the eventual symmetry
properties of the particular physical situation of interest. The
most trivial effect of this is the possibility of increasing the
numerical statistics of the results, but it also permits to force
the system to stay with a given symmetry, by eliminating from the
dynamics the degrees of freedom associated with the ``irrelevant''
directions. For instance, for our present purposes we want the
system to stay in the HCS, so that no spatial instabilities can be
developed in any direction, and at the same time we want to avoid
the introduction of finite size effects. This can be accomplished
by simulating the $N$-particle dynamics associated with the {\em
homogeneous} Boltzmann equation. The way of implementing this in
the simulation is by considering only one cell, i.e. any pair of
particles in the system can collide and no attention is paid to
their positions. Since the technical details of the method have
been extensively discussed in the literature \cite{Bi94,Ga00},
they will not be reproduced here.

The simulations we will present in the following correspond to a
two-dimensional system of hard disks, i.e. $d=2$ . As already
discussed, no boundary conditions are needed since we are
simulating homogeneous situations and the positions of the
particles play no role. For the same reason, there is no system
size to be specified. Moreover, the numerical data we will report
have been averaged over a number of trajectories, typically of the
order of a few thousands. The number of particles in the system is
$N=10^{4}$, but we insist on that it has only a statistical
meaning, since the dynamics of the system being used in the
simulation corresponds always by construction to that of a low
density gas.

Accordingly with the scenario we have developed, the simulation of
a low density gas whose underlying particle dynamics is defined by
Eqs. (\ref{3.3}) and (\ref{3.4}), is expected to yield a steady
state after a relatively short transient period. Averages of
properties in the steady state are simply related with the (time
dependent) properties of the HCS. In this way, the difficulties
associated with the fast cooling of the fluid when described in
terms of the actual variables, leading very soon to numerical
inaccuracies, are overcome. One technical point requiring some
attention is that the total momentum is unstable in the mapped
representation  for all sizes of the system \cite{Lu01}. Round-off
numerical errors lead to the presence of nonzero total momentum
that grow very quickly due to the instability. This is easily
avoid by calculating the total momentum at each simulation step
and subtracting it evenly from the momentum of each particle.

\begin{figure}
\includegraphics[scale=0.5,angle=0]{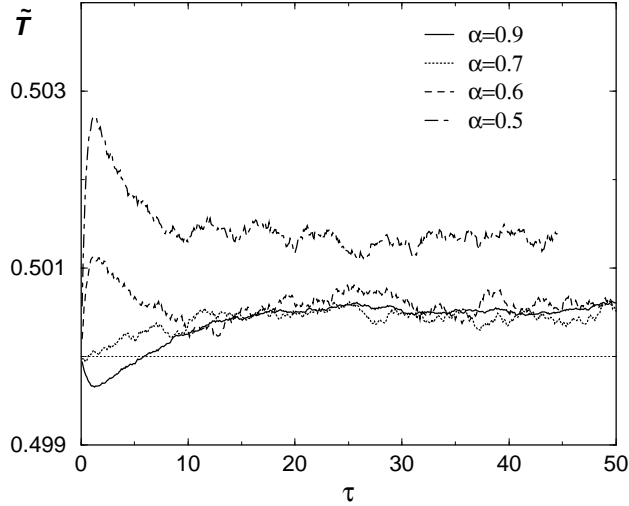}
\caption{\label{fig1} Evolution of the scaled temperature
$\widetilde{T}$ as a function of the scaled time $\tau$. Units are
defined in the main text. The different line styles correspond to
different values of $\alpha$ as indicated in the figure. }
\end{figure}

In the results to be reported in the following, the unit of time
is given by $\ell [2 k_{B}\widetilde{T}(0)/m]^{-1/2}$, the unit of
length is $\ell$, the unit of mass is $m$, and $k_{B}$ has been
set equal to unity. In Fig.\ \ref{fig1} the evolution of the
scaled temperature $\tilde{T}$ is plotted as a function of the
time $\tau$ for several values of $\alpha$, namely $\alpha=0.5$,
$0.6$, $0.7$, and $0.9$. The value of $\omega_{0}$ used in each
case is $\omega_{0}=\bar{\zeta} \widetilde{T}(0)^{1/2}/2$, with
$\bar{\zeta}(\alpha)$ approximated by its estimate in the first
Sonine approximation as given by Eqs. (\ref{2.14}) and
(\ref{2.23}). If this latter expression were exact, the final
steady temperature would be exactly the same as the initial one.
The observed behavior is consistent with the theoretical
predictions. In all the cases the temperature fluctuates around a
steady value after and initial transient time. In the original
time scale, this corresponds to the regime in which the granular
temperature has already reached its asymptotic form given in Eq.\
(\ref{2.13}). The steady value is very close to the initial
temperature, as expected, although a small deviation is clearly
identified for $\alpha=0.5$, indicating the approximated nature of
expression (\ref{2.23}) for $\zeta_{0}$. It is also seen in the
figure that the amplitude of the long time temperature
fluctuations increases as the value of the coefficient of
restitution decreases, i.e. as the system is more inelastic. This
is due to the change of the shape of the velocity distribution
function and also to the presence of velocity correlations in the
HCS that are generated by the $N$-particle dynamics of the system,
even in the low density limit. This will be discussed in more
detail elsewhere.

\begin{figure}
\includegraphics[scale=0.5,angle=0]{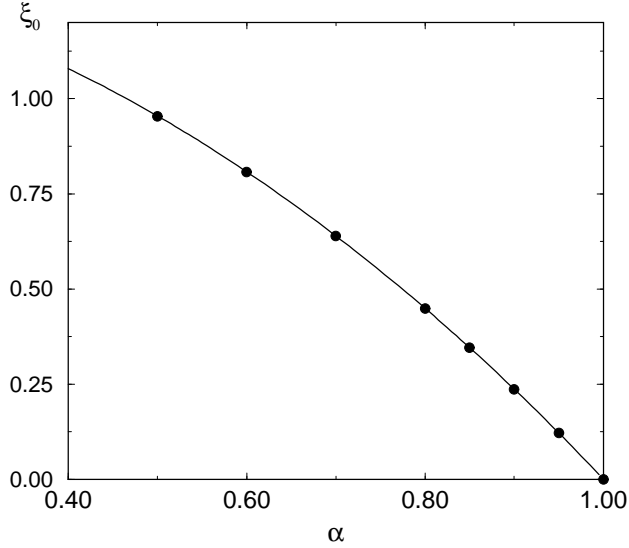}
\caption{\label{fig2} The dimensionless cooling rate $\zeta_{0}$
as a function of the coefficient of restitution $\alpha$. The
symbols are the values obtained from the simulations, while the
solid line is the theoretical prediction in the first Sonine
approximation.}
\end{figure}

>From the values of $\widetilde{T}_{st}$, the cooling rate
$\zeta_{0}$ can be obtained as a function of $\alpha$ by means of
Eq.\ (\ref{3.11}). In Fig. \ref{fig2} these numerical results are
compared with the theoretical prediction given by Eq.\
(\ref{2.23}). As already indicated by the weak dependence on
$\alpha$ of the steady temperature in Fig.\ \ref{fig1}, there is a
very good agreement. The discrepancy observed in  the latter for
$\alpha=0.5$ can not be made out on the scale used in Fig.
\ref{fig2}.

\begin{figure}
\includegraphics[scale=0.5,angle=0]{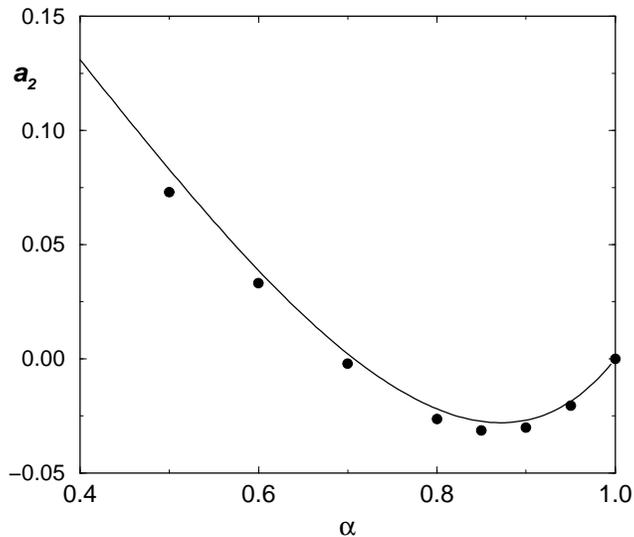}
\caption{\label{fig3} Coefficient $a_{2}$ defined in Eq.\
(\ref{2.20}) as a function of the coefficient of restitution
$\alpha$. The symbols are from the DSMC simulations, while the
solid line is the theoretical prediction discussed in the main
text.}
\end{figure}

Since the steady velocity distribution has the same form as
$\chi_{HCS}$, the steady representation also provides very
accurate data for it. As an example, in Fig. \ref{fig3} the
coefficient $a_{2}$ defined by Eq. (\ref{2.20}) is plotted as a
function of the coefficient of restitution and compared with the
approximated expression given by Eq.\ (\ref{2.22}). In this case,
a small but systematic deviation is observed. The results
presented so far in the above three figures are consistent, and
physically equivalent, to those obtained previously by carrying
out DSMC simulations in the actual variables of the system, i.e.
by dealing directly with the time-dependent HCS \cite{BRyC96}. The
main advantage of the present representation is that it reduces
the statistical uncertainties by introducing a steady state that
maps exactly into the HCS. In the next Section, it will be shown
that the scaling is also useful to study time correlation
functions.

\section{Averages and time-correlation functions in the steady
representation} \label{s4}

Consider a phase function of the form
\begin{equation}
\label{4.1} A(\Gamma)=\sum_{i=1}^{N} a({\bm R}_{i},{\bm V}_{i}),
\end{equation}
where $\Gamma$ denotes a point in the phase space of the system
and $a({\bm R}_{i},{\bm V}_{i})$ is a given one-particle dynamical
variable. The average value of $A$ at time $t$ is
\begin{equation}
\label{4.2} \langle A;t \rangle \equiv \int d\Gamma\, A(\Gamma,t)
\rho(\Gamma,0) \equiv \int d\Gamma\, A(\Gamma) \rho (\Gamma,t).
\end{equation}
Here $\rho(\Gamma,t)$ is the $N$-particle distribution function of
the system at time $t$. This equation is equivalent to
\begin{equation}
\label{4.3} \langle A;t \rangle = \int d{\bm r}\, \int d{\bm v}\,
a({\bm r},{\bm v}) f({\bm r},{\bm v},t),
\end{equation}
that, using the scaling defined by Eqs. (\ref{3.1}) and
(\ref{3.6}), can also be expressed as
\begin{equation}
\label{4.4} \langle A;t \rangle = \int d{\bm r}\, \int d{\bm w}\,
a \left[ {\bm r}, (\omega_{0} t)^{-1} {\bm w} \right] \tilde{f}
({\bm r},{\bm w}, \tau).
\end{equation}

Suppose now a dilute system in the HCS. It has been established in
Sec. \ref{s3} that in the long time limit, the scaled distribution
function tends to the steady form $\tilde{f}_{st}({\bm w})$ given
by Eq.\ (\ref{3.13}), so that Eq.\ (\ref{4.4}) becomes
\begin{equation}
\label{4.5} \langle A;t \rangle_{HCS}=\int d{\bm r}\, \int d{\bm
w}\, a\left[ {\bm r}, \frac{v_{0}(t)}{\widetilde{v}_{0,st}}\, {\bm
w} \right] \tilde{f}_{st}({\bm w}).
\end{equation}
This is the low density limit of
\begin{equation}
\label{4.6} \langle A;t \rangle_{HCS}=\int d \widetilde{\Gamma}\,
\widetilde{A}(\widetilde{\Gamma})
\tilde{\rho}_{st}(\widetilde{\Gamma}),
\end{equation}
where $\widetilde{\Gamma}=\{ {\bm R}_{i},{\bm W}_{i}; i=1,\cdots,
N \}$,
\begin{equation}
\label{4.7}
\widetilde{A}(\widetilde{\Gamma})=\sum_{i=1}^{N}a\left[{\bm
R}_{i},\frac{v_{0}(t)}{\widetilde{v}_{0,st}} {\bm W}_{i} \right],
\end{equation}
and $\tilde{\rho}_{st}(\widetilde{\Gamma})$ is the scaled
$N$-particle distribution of the steady state.

In this way, averages in the HCS are exactly mapped onto averages
in the the steady state of the scaled dynamics. It could be
thought that since Eq.\ (\ref{4.5}) has been obtained as the long
time limit of Eq.\ (\ref{4.4}) for the HCS, its validity is
somewhat restricted to very low actual temperatures $T_{HCS}(t)$.
Nevertheless, this is not the case, because formally the initial
temperature $T_{HCS}(0)$ can be as large as wanted and,
consequently, Eq.\ (\ref{4.5}) can be applied at any temperature
$T_{HCS}(t)$. In fact, it is easily verified that the same
equation can be derived from the particularization of Eq.\
({\ref{4.3}) to the HCS by making the change of variable ${\bm
w}=\widetilde{v}_{0,st} {\bm c}$, without introducing any long
time limit. Of course, the difference is that in this latter case
$T_{HCS}(t)$, present in Eq.\, (\ref{4.5}) through $v_{0}(t)$, can
not be substituted by its asymptotic form.

Next, let us consider time correlation functions in the HCS
defined by
\begin{equation}
\label{4.8} C_{AB}(t,t^{\prime}) \equiv \langle A(t)
B(t^{\prime});0 \rangle_{HCS}-\langle A;t \rangle_{HCS} \langle
B;t^{\prime} \rangle_{HCS},
\end{equation}
with
\begin{equation}
\label{4.9} \langle A(t) B(t^{\prime});0 \rangle_{HCS} \equiv \int
d \Gamma\, A(\Gamma,t) B(\Gamma,t^{\prime}) \rho_{HCS} (\Gamma,0)
\end{equation}
and it is assumed that $t>t^{\prime}>0$. The phase functions
$A(\Gamma)$ and $B(\Gamma)$ are the sum of one-particle dynamical
variables, i.e. $A$  is again  given by Eq.\ (\ref{4.1}) and
\begin{equation}
\label{4.10} B(\Gamma)=\sum_{i=1}^{N} b({\bm R}_{i},{\bm V}_{i}).
\end{equation}
Then, Eq.\ (\ref{4.8}) can be rewritten in the equivalent form
\begin{equation}
\label{4.11} C_{AB}(t,t^{\prime})=\int d{\bm r}_{1} \int d{\bm
v}_{1} \int d{\bm r}^{\prime}_{1} \int d{\bm v}^{\prime}_{1}\,
a({\bm r}_{1},{\bm v}_{1})b({\bm r}^{\prime}_{1},{\bm
v}^{\prime}_{1}) h_{1/1}({\bm r}_{1},{\bm v}_{1},t; {\bm
r}_{1}^{\prime}, {\bm v}_{1}^{\prime},t^{\prime}),
\end{equation}
where $h_{1/1}$ is the two-particle two-time correlation function
of the HCS defined by \cite{EyC81}
\begin{equation}
\label{4.12} h_{1/1}({\bm r}_{1},{\bm v}_{1},t; {\bm
r}_{1}^{\prime}, {\bm v}_{1}^{\prime},t^{\prime}) \equiv
f_{1/1}({\bm r}_{1},{\bm v}_{1},t; {\bm r}_{1}^{\prime}, {\bm
v}_{1}^{\prime},t^{\prime}) -f_{HCS} ({\bm v}_{1},t) f_{HCS}({\bm
v}_{1}^{\prime},t).
\end{equation}
Here $f_{1/1}$ is the two-particle two-time reduced distribution
function of the HCS,
\begin{eqnarray}
\label{4.13}  f_{1/1}({\bm r}_{1},{\bm v}_{1},t ; {\bm
r}_{1}^{\prime}, {\bm v}_{1}^{\prime},t^{\prime})  & \equiv &
\sum_{i=1}^{N} \sum_{j=1}^{N} \int d\Gamma\, \rho_{HCS}(\Gamma)
\delta [{\bm r}_{1}-{\bm R}_{i}(t)]\, \delta[{\bm v}_{1}-{\bm
V}_{i}(t)]\, \nonumber \\
& & \times \delta [{\bm r}_{1}^{\prime}-{\bm R}_{j}(t)]\,
\delta[{\bm v}_{1}^{\prime}-{\bm V}_{j}(t)].
\end{eqnarray}
In the low density limit, and by using the same kind of
assumptions as needed to derive the Boltzmann equation, it is
possible to obtain a formal expression for $h_{1/1}$ involving
only the steady one-particle distribution function \cite{EyC81}. A
sketch of the calculations is given in the Appendix \ref{ap1}.
When this expression is substituted into Eq.\ (\ref{4.11}) it is
obtained:
\begin{equation}
\label{4.14} C_{AB}(t,t^{\prime})= \int d{\bm r}_{1} \int d{\bm
w}_{1}\, \tilde{f}_{st}({\bm w}_{1}) a \left[{\bm
r}_{1},\frac{v_{0}(t)}{\widetilde{v}_{0,st}}\, {\bm
w}_{1},\tau-\tau_{1} \right] b \left[ {\bm r}_{1},
\frac{v_{0}(t^{\prime})}{\widetilde{v}_{0,st}}\, {\bm w}_{1}
\right].
\end{equation}
The time dependence of the dynamical variable $a$ is given by
\begin{equation}
\label{4.15} a({\bf r}_{1},{\bf w}_{1},\tau)=e^{\tau \bar{L}_{B}
({\bm w}_{1})} a({\bm r}_{1},{\bm w}_{1}).
\end{equation}
where $\bar{L}_{B}({\bm w}_{1})$ is the adjoint of the linearized
Boltzmann operator around the steady state in the scaled dynamics,
\begin{equation}
\label{4.16} \bar{L}_{B}({\bm w}_{1})= {\bm w}_{1} \cdot
\frac{\partial}{\partial {\bm r}_{1}} + \bar{\Lambda}_{st} ({\bm
w}_{1}),
\end{equation}
\begin{equation}
\label{4.17} \bar{\Lambda}_{st}=\omega_{0} {\bm w}_{1} \cdot
\frac{\partial}{\partial {\bm w}_{1}} +\bar{K}_{st} ({\bm w}_{1}),
\end{equation}
\begin{equation}
\label{4.18} \bar{K}_{st}({\bm w}_{1})= \sigma^{d-1} \int d {\bm
w}_{2}\, \tilde{f}_{st} ({\bm w}_{2}) \int d\widehat{\bm \sigma}\,
\Theta ({\bm w}_{12}\cdot \widehat{\bm \sigma}) {\bm w}_{12}\cdot
\widehat{\bm \sigma}[b_{\bm \sigma}({\bm w}_{1},{\bm w}_{2}) -1]
(1+\mathcal{P}_{12}).
\end{equation}
The operator $\mathcal{P}_{12}$ interchanges the subindexes $1$
and $2$ to its right. Of course, if the dynamical variable $a$
does not depend on the position ${\bm r}$, $\bar{L}_{B}({\bm
w}_{1})$ can be substituted by $\bar{\Lambda}_{st}({\bm w}_{1})$
in Eq.\ (\ref{4.15}).

In the particular, but quite frequent case that $a$ and $b$ are
homogeneous functions of the velocity of degree $q_{1}$ and
$q_{2}$, respectively, i.e.
\begin{eqnarray}
\label{4.19} a \left[ {\bm r}_{1},
\frac{v_{0}(t)}{\widetilde{v}_{0,st}}\, {\bm w}_{1} \right] &=&
\left[ \frac{v_{0}(t)}{\widetilde{v}_{0,st}} \right]^{q_{1}}
a({\bm r}_{1},{\bm w}_{1}), \nonumber \\
b \left[ {\bm r}_{1},
\frac{v_{0}(t^{\prime})}{\widetilde{v}_{0,st}}\, {\bm w}_{1}
\right] &=& \left[ \frac{v_{0}(t^{\prime})}{\widetilde{v}_{0,st}}
\right]^{q_{2}} b({\bm r}_{1},{\bm w}_{1}),
\end{eqnarray}
the time-correlation function becomes
\begin{equation}
\label{4.20} C_{AB}(t,t^{\prime})=\frac{v_{0}^{q_{1}}(t)
v_{0}^{q_{2}}(t^{\prime})}{\widetilde{v}_{0,st}^{q_{1}+q_{2}}}
\int d{\bm r}_{1} \int d{\bm w}_{1}\, \tilde{f}_{st}({\bm w}_{1})
a ({\bm r}_{1}, {\bm w}_{1},\tau-\tau_{1}) b( {\bm r}_{1},{\bm
w}_{1}).
\end{equation}

In the context of the steady representation of the HCS of a low
density granular gas as discussed in this paper, the relevant
point of the above analysis is the following: suppose that a
property of the gas can be expressed in terms of the time
correlation function
\begin{equation}
\label{4.21} \langle a(\tau) b \rangle_{st} \equiv \int d{\bm r}
\int d{\bm v}\, \tilde{f}_{st}({\bm v}) a({\bm r},{\bm v},\tau)
b({\bm r}, {\bm v}),
\end{equation}
with the time-dependence of $a({\bm r},{\bm v}, \tau)$ given by
Eq. (\ref{4.15}). A slight modification of the preceding
discussion shows that this is the low density limit of
\begin{equation}
\label{4.22} C_{AB,st}(\tau)=\langle A(\tau) B;0 \rangle_{st}
-\langle A \rangle_{st} \langle B \rangle_{st},
\end{equation}
where
\begin{equation}
\label{4.23} \langle A \rangle_{st} =\int d\Gamma\,
\rho_{st}(\Gamma) A (\Gamma), \quad \langle B \rangle_{st}= \int
d\Gamma\, \rho_{st}(\Gamma)B(\Gamma),
\end{equation}
\begin{equation}
\label{4.24} \langle A(\tau) B;0 \rangle_{st}= \int d\Gamma\,
\rho_{st}(\Gamma) A (\Gamma,\tau) B(\Gamma),
\end{equation}
with $A(\Gamma)$ and $B(\Gamma)$ given by Eqs.\ (\ref{4.1}) and
(\ref{4.10}), respectively.  Besides, $A(\Gamma,\tau)$ is
generated from $A(\Gamma)$ by the dynamics defined by Eqs.
(\ref{3.3}) and (\ref{3.4}), and $\rho_{st}(\Gamma)$ is the
$N$-particle distribution function of the steady state eventually
reached by the system. Although the existence of this state is not
rigorously proven, it is supported by the discussion carried in
Sec.\ \ref{s3} at the level of the Boltzmann equation and also by
Molecular Dynamics simulations \cite{Lu01,LByD02}. The DSMC method
provides an efficient tool to generate the $N$-particle dynamical
representation of a dilute gas, that is consistent with the
Boltzmann equation. In summary, it allows the direct evaluation of
Eq.\ (\ref{4.22}) in the low density limit, where it is equivalent
to Eq.\ (\ref{4.21}).

An important application of the above ideas is the evaluation of
the Navier-Stokes transport coefficients of a dilute granular gas
composed of hard spheres or disks. In refs.\
\cite{DyB02,DyB03,BDyR03}, these coefficients were derived from
the inelastic Boltzmann equation by means of the Chapman-Enskog
procedure, eigenfunctions expansions, and linear response theory,
finding equivalent results. The expressions of all the transport
coefficients are proportional to time integrals of correlation
functions of the form
\begin{equation}
\label{4.25} D_{ab}(s)=\int d{\bm c}_{1}\, \chi_{HCS}({\bm c}_{1})
a({\bm c}_{1},s) b({\bm c}_{1}),
\end{equation}
with the time dependence of $a({\bm c},s)$ determined by
\begin{equation}
\label{4.26} a({\bm c}_{1},s)= e^{s\bar{\Lambda}_{c}({\bm c}_{1})}
a({\bm c}_{1}),
\end{equation}
\begin{equation}
\label{4.27} \bar{\Lambda}_{c}({\bm c}_{1})= \int d {\bm c}_{2}\,
\chi_{HCS} ({\bm c}_{2}) \int d\widehat{\bm \sigma}\, \Theta ({\bm
c}_{12}\cdot \widehat{\bm \sigma}) {\bm c}_{12}\cdot \widehat{\bm
\sigma}[b_{\bm \sigma}({\bm c}_{1},{\bm c}_{2})-1]
(1+\mathcal{P}_{12})+ \frac{\zeta_{0}}{2} {\bm c}_{1} \cdot
\frac{\partial}{\partial {\bm c}_{1}}.
\end{equation}
Finally, the time scale $s$ is defined as
\begin{equation}
\label{4.28} s=\int_{0}^{t} dt^{\prime}\,
\frac{v_{0}(t^{\prime})}{\ell}\, .
\end{equation}

Although the above representation is appropriate for formal
manipulations, for computational purposes the dynamics associated
with the operator $\bar{\Lambda}_{c}$ presents the technical
complication that it involves the cooling rate $\zeta_{0}$, that
therefore must be known a priori instead of being determined by
the own simulation, as it is the case for the scaled dynamics
being used in the present paper. For this reason, it is convenient
to transform Eq.\ (\ref{4.25}) by writing it in terms of that
dynamics. This is easily accomplished by introducing
\begin{equation}
\label{4.29} \tau=\frac{\zeta_{0}}{2 \omega_{0}} s, \quad {\bm
w}=\frac{2 \omega_{0} \ell}{\zeta_{0}}  {\bm c}
=\widetilde{v}_{0,st} {\bm c}.
\end{equation}
Then, for an arbitrary function $g({\bm c})$ it is obtained
\begin{equation}
\label{4.30} s \bar{\Lambda}_{c} g({\bm c})=\tau
\bar{\Lambda}_{st} g \left( \frac{\bm w}{\widetilde{v}_{0,st}}
\right),
\end{equation}
where $\bar{\Lambda}_{st}$ is the operator defined by Eq.\
(\ref{4.17}). It follows that Eq.\ (\ref{4.25}) is the same as
\begin{equation}
\label{4.31} D_{ab}(s)= \frac{1}{N} \langle a \left( \frac{\bm
v}{\widetilde{v}_{0,st}}\, ,\tau \right) b \left( \frac{\bm v}{
\widetilde{v}_{0,st}} \right) \rangle_{st},
\end{equation}
where  we use the notation introduced in Eq.\, (\ref{4.21}). The
above result relates the expression of the transport coefficients
of a dilute granular gas, as obtained from the Boltzmann equation
for the one-particle distribution function, with the low density
limit of time-correlation functions in the $N$-particle dynamics,
computed in the steady state of the scaled representation
introduced in Sec.\ \ref{s3}.

\section{Self-diffusion}
\label{s5} In ref. \cite{BRCyG00}, the expression of the
self-diffusion coefficient $D$ of a dilute inelastic gas of hard
particles has been derived from the Boltzmann-Lorentz equation by
the Chapman-Enskog procedure. In Appendix \ref{ap2} it is shown
that the results reported in \cite{BRCyG00} can be expressed in
the form
\begin{eqnarray}
\label{5.1} D & = & \frac{v_{0}(t) \ell}{d} \int_{0}^{\infty} ds
\int d{\bm c}\, \chi_{HCS}({\bm c}) {\bm c}(s) \cdot {\bm c}
\nonumber \\
& = & \frac{v_{0}(t)}{\widetilde{v}_{0,st} N d} \int_{0}^{\infty}
\langle {\bm w}(t) \cdot {\bm w} \rangle_{st}.
\end{eqnarray}
In the last transformation we have used Eq.\ (\ref{4.31}). In
fact, this is an special case, since the time dependence of ${\bm
c}(s)$ is not given by the linearized Boltzmann collision operator
$\bar{\Lambda}_{\bm c}$ as in Eq.\ (\ref{4.26}), but by the
Lorentz-Boltzmann one $\bar{\Lambda}_{BL,c}({\bf c})$, defined in
Eq.\ (\ref{b11}),
\begin{equation}
\label{5.2} {\bm c}(s)= e^{s \bar{\Lambda}_{BL,c}({\bm c})} {\bm
c}.
\end{equation}
Consequently, in Eq.\ (\ref{5.1}) it is
\begin{equation}
\label{5.3} {\bm w}(\tau)= e^{\tau  \bar{\Lambda}_{BL,st}({\bm
w})} {\bm w},
\end{equation}
\begin{equation}
\label{5.4} \bar{\Lambda}_{BL,st}({\bm w})= \sigma^{d-1} \int d
{\bm w}_{1}\, \tilde{f}_{st} ({\bm w}_{1}) \int d\widehat{\bm
\sigma}\, \Theta ({\bm g}\cdot \widehat{\bm \sigma}) {\bm g} \cdot
\widehat{\bm \sigma}[b_{\bm \sigma}({\bm w},{\bm w}_{1})-1] +
\frac{\zeta_{0}}{2} {\bm w} \cdot \frac{\partial}{\partial {\bm
w}}.
\end{equation}
Nevertheless, all the reasonings in Sec.\ \ref{s4} and Appendix
\ref{ap1} can be easily adapted to the present case. The only, but
relevant, difference, is that now the expression $\langle {\bm
w}(t) \cdot {\bm w} \rangle_{st}$ is the low density limit of
\begin{equation}
\label{5.5} C_{vv}(\tau) \equiv \sum_{i=1}^{N} \int d \Gamma\,
\rho_{st}(\Gamma) {\bm W}_{i}(t) \cdot {\bm W}_{i}
\end{equation}
and not of
\begin{equation}
\label{5.6} \sum_{i=1}^{N}\sum_{j=1}^{N} \int d \Gamma\,
\rho_{st}(\Gamma) {\bm W}_{i}(t) \cdot {\bm W}_{j}.
\end{equation}
Of course, this has to be taken into account when computing the
self-diffusion coefficient by means of DSMC simulations.

Figure \ref{fig4} shows the velocity autocorrelation function
$C_{vv}(\tau)$ for three different values of the coefficient of
restitution. It is seen that it decays quite fast to zero,
implying the convergence of its time integral and, therefore, the
existence of the constant diffusion coefficient. The value of the
velocity autocorrelation at each time $\tau$ is based on an
average of all the times included in the simulation, once the
steady state has been reached. In Fig.\ \ref{fig5}, the same
quantity is plotted on a logarithmic scale where an exponential
decay for all times is clearly identified.

\begin{figure}
\includegraphics[scale=0.5,angle=0]{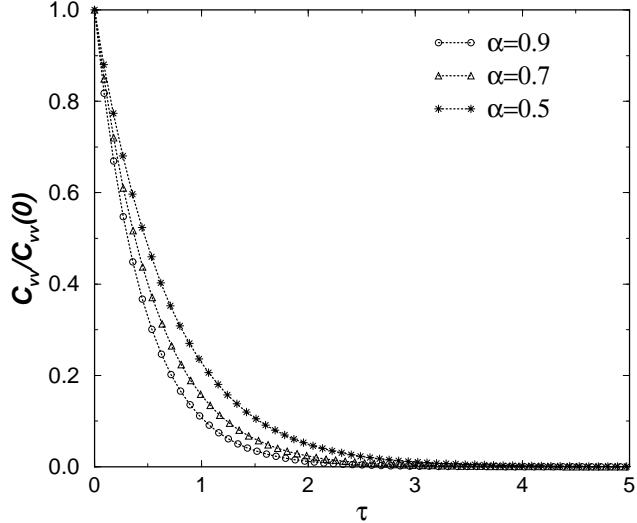}
\caption{\label{fig4} Time evolution of the normalized velocity
autocorrelation function $C_{vv}(\tau)/C_{vv}(0)$ as a function of
the scaled time $\tau$. The results for three different values of
the coefficient of restitution $\alpha$ are shown, as indicated in
the figure. }
\end{figure}

\begin{figure}
\includegraphics[scale=0.5,angle=0]{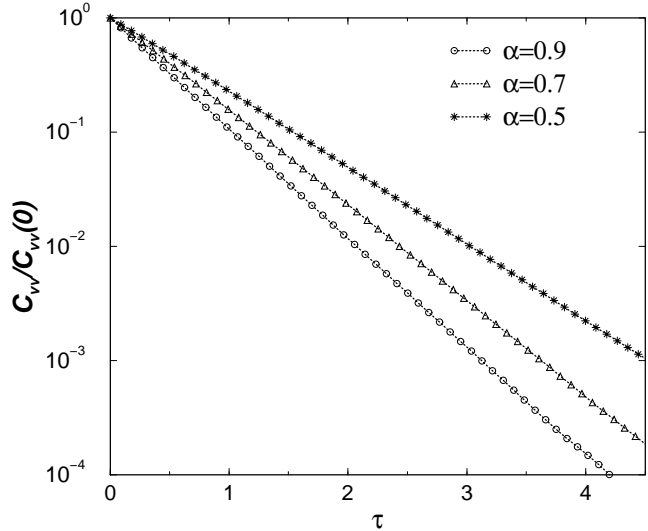}
\caption{\label{fig5} The same as in Fig. \ref{fig4} but on a
logarithmic scale.}
\end{figure}

>From the velocity autocorrelation function the self-diffusion
coefficient is obtained by means of the Green-Kubo relation
(\ref{5.1}). Figure \ref{fig6} shows the ratio of the obtained
values $D(T_{HCS},\alpha)$ to the elastic limit ($\alpha
\rightarrow 1$) of the value predicted by the Chapman-Enskog
method in the first Sonine approximation $D_{0}(T_{HCS})$, namely
\begin{equation}
\label{5.7} D^{*}(\alpha) \equiv
\frac{D(T_{HCS},\alpha)}{D_{0}(T_{HCS})},
\end{equation}
\begin{equation}
\label{5.8} D_{0}(T_{HCS})= \frac{\Gamma(d/2) d}{ 4 \pi^{(d-1)/2}
n \sigma^{d-1}} \left( \frac{k_{B}T_{HCS}}{m} \right)^{1/2}.
\end{equation}
Also plotted is the theoretical prediction for $D^{*}(\alpha)$
again in the first Sonine approximation \cite{BRCyG00},
\begin{equation}
\label{5.9} D^{*}(\alpha)=4 \left[ (1+\alpha)^{2} -
\frac{a_{2}(\alpha)}{16}(4+\alpha-3 \alpha^{2}) \right]^{-1}.
\end{equation}

The agreement between theory and simulation is quite good in all
the range of values of $\alpha$ considered, although a systematic
deviation, which increases as the value of $\alpha$ decreases, is
observed. Self-diffusion in the HCS  of a system of inelastic hard
spheres ($d=3$) has been previously studied by means of DSMC in
the actual variables, i.e. under continuous cooling, in ref.
\cite{BRCyG00}, where the diffusion constant was calculated from
the mean-square displacement by means of the (inelastic) Einstein
relation and also from the linear response of the system to a
density perturbation. The comparison between the numerical results
obtained there and the theoretical prediction given by Eq.\
(\ref{5.9}) is very similar to the one presented in Fig.\
\ref{fig6}. Very recently \cite{GyM03}, it has been shown that the
agreement improves over the whole range of values of $\alpha$ if
$D^{*}(\alpha)$ is computed in the second Sonine approximation.
Although the analysis is restricted to the case of inelastic hard
spheres, it seems sensible to expect the same kind of behavior for
a system of inelastic hard disks.

\begin{figure}
\includegraphics[scale=0.5,angle=0]{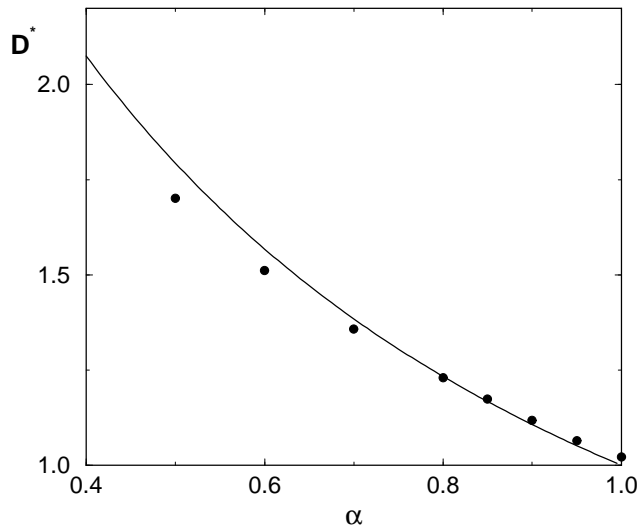}
\caption{\label{fig6} Dimensionless reduced diffusion coefficient
$D^{*}$ as a function of the coefficient of restitution $\alpha$.
The symbols are from the simulations, while the solid line is the
theoretical prediction from the Chapman-Enskog procedure in the
first Sonine approximation. }
\end{figure}

As already mentioned several times, the main feature of the
numerical method used here is that it takes advantage of the
steady representation of the HCS, removing any limitation on the
time on which trajectories of the system may be followed. In
addition, let us point out that the analysis of the self-diffusion
coefficient based on the velocity autocorrelation function as
developed here, is expected to give more accurate results, from a
statistical point of view, than those based on the mean square
displacement. This is because, while each value of the former in a
given trajectory of the system is obtained from an average over
all the simulation interval, each value of the latter is obtained
from a single evaluation.

\section{Discussion and conclusion}
\label{s6} In this paper, the actual dynamics of a low density
granular gas composed of smooth hard inelastic particles has been
transformed to another one in which the HCS becomes
time-independent. In this way, the need for more or less
uncontrolled mechanisms such as internal thermostats in order to
get a stationary state is eliminated. The transformation is
closely related to the fact that the temperature of the HCS
becomes independent from its initial value in the asymptotically
long time limit, a property that has not received enough attention
up to now.  The exact correspondence between both formulations for
averages and time correlation functions has been explicitly
established. This requires to consider the kinetic equation for
the one-particle distribution function, i.e.  the inelastic
Boltzmann equation, and also the equation for the two-particle and
two-time correlation function in the low density limit. The latter
is obtained by extending in a natural way the standard methods of
kinetic theory.  In particular, it has been shown that the steady
temperature directly determines the value of the cooling rate of
the system. Then, the DSMC method has been used to measure it and
the numerical results have been compared with the theoretical
predictions obtained by solving the Boltzmann equation in the
first Sonine approximation.

The introduction of the steady representation of the HCS enables
the evaluation of the Green-Kubo expressions for the transport
coefficients of a dilute granular gas by means of the DSMC method,
just as for normal fluids whose particles collide elastically. To
put this in a proper context, we have emphasized that the DSMC
method is not just a numerical tool to solve the Boltzmann
equation, but it formulates an effective N-body dynamics that is
expected to be equivalent to the Newtonian one in the low density
limit. Moreover, it has the advantage of allowing to incorporate
in the own effective dynamics of the particles the symmetry
properties of the particular state being simulated. So, it is
possible to force the system to stay homogeneous, avoiding the
spontaneous development of spacial inhomogeneities associated with
the clustering instability. Here this has been illustrated for the
simplest case of self-diffusion, whose expression involves the
velocity autocorrelation function. The results have been compared
with the theoretical prediction obtained by the Chapman-Enskog
method and also with  previous numerical simulations carried out
in the actual dynamics in which the HCS is time-dependent. The
study of the remaining Navier-Stokes transport coefficients will
be reported elsewhere.

In summary, we believe that the transformed dynamics discussed in
this work is of both formal and practical interest for the study
of granular fluids in the low density limit. Although we have
restricted here ourselves to its application to the HCS, the
mapping with the actual dynamics can be easily extended to any
arbitrary state.
\begin{acknowledgments}
We acknowledge  economical support from the Ministerio de Ciencia
y Tecnolog\'{\i}a (Spain) through Grant No. BFM2002-00303
(partially financed by FEDER funds)
\end{acknowledgments}

\appendix

\section{Low density limit of time-correlation functions in the HCS}
\label{ap1} In the low density limit, neglecting three-particle
correlations, the function $h_{1/1}$ obeys the equation,
\begin{equation}
\label{a1} \left[ \frac{\partial}{\partial t}+{\bm v}_{1} \cdot
\frac{\partial}{\partial {\bm r}_{1}} -K({\bm v}_{1},t) \right]
h_{1/1}({\bm r}_{1},{\bm v}_{1},t; {\bm r}_{1}^{\prime}, {\bm
v}_{1}^{\prime},t^{\prime})=0,
\end{equation}
where $K$ is the linearized Bolztmann operator
\begin{equation}
\label{a2} K({\bm v}_{1},t)=\sigma^{d-1} \int d{\bm v}_{2} \int d
\widehat{\bm \sigma}\, \Theta ({\bm v}_{12} \cdot \widehat{\bm
\sigma}) |{\bm v}_{12} \cdot \widehat{\bm \sigma}| (\alpha^{-2}
b_{\bm \sigma}-1) (1+\mathcal{P}_{12}) f_{HCS}({\bm v}_{2},t).
\end{equation}
The permutation operator $\mathcal{P}_{12}$ interchanges the
indexes $1$ and $2$ of the velocities appearing to its right. The
above equation holds for $t>t^{\prime}>0$ and it has to be solved
with the initial condition
\begin{equation}
\label{a3} h_{1/1}({\bm r}_{1},{\bm v}_{1},t^{\prime}; {\bm
r}_{1}^{\prime}, {\bm v}_{1}^{\prime},t^{\prime})=g_{2,HCS}({\bm
r}_{1},{\bm v}_{1},{\bm r}_{1}^{\prime},{\bm
v}_{1}^{\prime},t^{\prime})+ \delta ({\bm r}_{1}-{\bm
r}^{\prime}_{1})\, \delta ({\bm v}_{1}-{\bm v}_{1}^{\prime})
f_{HCS}({\bm v}_{1},t^{\prime}),
\end{equation}
$g_{2,HCS}$ being the two-particle one-time correlation function
of the HCS. Equation (\ref{a1}) can be derived by the hierarchy
method starting from the pseudo-Liouville equation for a system of
inelastic hard spheres or disks. A detailed analysis for the
elastic case is presented in ref. \cite{EyC81}. Since the same
method can be applied {\em mutatis mutandis} to the inelastic
case, it will be not reproduced here. Now, time and velocities are
scaled by
\begin{equation}
\label{a4} \tau =\int_{0}^{t} dt^{\prime}\,
\frac{v_{0}(t^{\prime})}{\widetilde{v}_{0,st}}
\end{equation}
and
\begin{equation}
\label{a5} {\bm w}=\frac{\widetilde{v}_{0,st}}{v_{0}(t)}\, {\bm
v},
\end{equation}
respectively. We are using the same symbols as in Sec. \ref{s3}
since we have seen that in the long time limit the scaling defined
by Eq.\ (\ref{a4}) is equivalent to that defined by Eq.\
(\ref{3.1}). The scaled correlation function is
\begin{equation}
\label{a6} \tilde{h}_{1/1}({\bm r}_{1},{\bm w}_{1},\tau;{\bm
r}_{1}^{\prime},{\bm w}_{1}^{\prime},\tau^{\prime})= \left[
\frac{v_{0}(t) v_{0}(t^{\prime})}{\widetilde{v}_{0,st}^{2}}
\right]^{d} h_{1/1}({\bm r}_{1},{\bm v}_{1},t; {\bm
r}_{1}^{\prime}, {\bm v}_{1}^{\prime},t^{\prime})
\end{equation}
and, in terms of the new variables, Eq.\ (\ref{a1}) becomes
\begin{equation}
\label{a7} \left[ \frac{\partial}{\partial \tau} +{\bm w}_{1}
\cdot \frac{\partial}{\partial {\bm r}_{1}} -\Lambda_{st} ({\bm
w}_{1}) \right]\tilde{h}_{1/1}({\bm r}_{1},{\bm w}_{1},\tau;{\bm
r}_{1}^{\prime},{\bm w}_{1}^{\prime},\tau^{\prime})=0,
\end{equation}
valid for $\tau> \tau^{\prime}$. The operator $\Lambda_{st}$ is
defined by
\begin{equation}
\label{a8} \Lambda_{st}({\bm w}_{1})= K_{st}({\bm
w}_{1})-\omega_{0} \frac{\partial}{\partial {\bm w}_{1}} \cdot
{\bm w}_{1},
\end{equation}
where $K_{st}$ is the steady linear Boltzmann collision operator
\begin{equation}
\label{a9} K_{st}({\bm w}_{1})=\sigma^{d-1} \int d{\bm w}_{2} \int
d\widehat{\bm \sigma}\, \Theta ({\bm w}_{12} \cdot \widehat{\bm
\sigma}) {\bm w}_{12} \cdot \widehat{\bm \sigma} \left[
\alpha^{-2} b_{\bm \sigma}^{-1} ({\bm w}_{1},{\bm w}_{2})-1
\right] (1+\mathcal{P}_{12}) \tilde{f}_{st}({\bm w}_{2}).
\end{equation}
Integration of Eq.\ (\ref{a7}) gives
\begin{equation}
\label{a10} \tilde{h}_{1/1}({\bm r}_{1},{\bm w}_{1},\tau;{\bm
r}_{1}^{\prime},{\bm w}_{1}^{\prime},\tau^{\prime})= e^{(\tau -
\tau^{\prime})L_{B}({\bm w}_{1})} \tilde{h}_{1/1}({\bm r}_{1},{\bm
w}_{1},\tau^{\prime};{\bm r}_{1}^{\prime},{\bm
w}_{1}^{\prime},\tau^{\prime}),
\end{equation}
\begin{equation}
\label{a11} L_{B}({\bm w}_{1})=\Lambda_{st}({\bm w}_{1})-{\bm
w}_{1} \cdot \frac{\partial}{\partial {\bm r}_{1}}.
\end{equation}
The initial condition on the right hand side is
\begin{equation}
\label{a12} \tilde{h}_{1/1}({\bm r}_{1},{\bm
w}_{1},\tau^{\prime};{\bm r}_{1}^{\prime},{\bm
w}_{1}^{\prime},\tau^{\prime}) =\tilde{g}_{2,HCS} ({\bm
r}_{1},{\bm w}_{1},{\bm r}^{\prime}_{1},{\bm
w}^{\prime}_{1},\tau^{\prime})+\delta ({\bm r}_{1}-{\bm
r}^{\prime}_{1})\, \delta ({\bm w}_{1}-{\bm w}^{\prime}_{1})
\tilde{f}_{st}({\bm w}_{1}),
\end{equation}
with
\begin{equation}
\label{a13} \tilde{g}_{2,HCS} ({\bm r}_{1},{\bm w}_{1},{\bm
r}^{\prime}_{1},{\bm w}^{\prime}_{1},\tau^{\prime})=\left[
\frac{v_{0}(t^{\prime})}{\widetilde{v}_{0,st}} \right]^{2d}
g_{2}({\bm r}_{1},{\bm v}_{1},{\bm r}^{\prime}_{1},{\bm
v}^{\prime}_{1},t^{\prime}).
\end{equation}

Now the assumption is made that the contribution to Eq.\
(\ref{a10}) coming from $\tilde{g}_{2}$ can be neglected in the
low density limit we are considering. Although there is no
explicit proof for this, it is consistent with the hypothesis made
to derive the Boltzmann equation. On the other hand, it must be
realized that the HCS is not an equilibrium state and, therefore,
it can present relevant position and velocity correlations, but we
expect them to become negligible for asymptotically small
densities. Then, substitution of Eq. (\ref{a12}) into Eq.\
(\ref{4.11}) yields
\begin{eqnarray}
\label{a14} C_{AB}(t,t^{\prime})&=& \int d{\bm r}_{1} \int d{\bm
w}_{1} \int d{\bm r}^{\prime}_{1} \int d{\bm w}^{\prime}_{1}\, a
\left[ {\bm r}_{1}, \frac{v_{0}(t)}{\widetilde{v}_{0,st}}\, {\bm
w}_{1} \right] b \left[ {\bm r}^{\prime}_{1},
\frac{v_{0}(t^{\prime})}{\widetilde{v}_{0,st}} {\bm
w}^{\prime}_{1} \right] \nonumber
\\ && e^{(\tau-\tau^{\prime})L_{B}({\bm w}_{1})} \delta({\bm
r_{1}}-{\bm r}^{\prime}_{1})\, \delta ({\bm w}_{1}-{\bm
w}^{\prime}_{1}) \tilde{f}_{st}({\bm w}_{1}) \nonumber \\
&=& \int d{\bm r}_{1} \int d{\bm w}_{1} \int d{\bm r}^{\prime}_{1}
\int d{\bm w}^{\prime}_{1}\ b \left[ {\bm r}^{\prime}_{1},
\frac{v_{0}(t^{\prime})}{\widetilde{v}_{0,st}} {\bm
w}^{\prime}_{1} \right] \delta({\bm r_{1}}-{\bm r}^{\prime}_{1})\,
\delta ({\bm w}_{1}-{\bm w}^{\prime}_{1}) \tilde{f}_{st}({\bm
w}_{1}) \nonumber \\ && e^{(\tau-\tau^{\prime})\bar{L}_{B}({\bm
w}_{1})} a \left[ {\bm r}_{1},
\frac{v_{0}(t)}{\widetilde{v}_{0,st}}\, {\bm w}_{1} \right]
\nonumber \\
&=& \int d{\bm r}_{1} \int d{\bm w}_{1}\, \tilde{f}_{st}({\bm
w}_{1}) b \left[ {\bm r}_{1},
\frac{v_{0}(t^{\prime})}{\widetilde{v}_{0,st}} {\bm w}_{1}
\right]e^{(\tau-\tau^{\prime})\bar{L}_{B}({\bm w}_{1})} a \left[
{\bm r}_{1}, \frac{v_{0}(t)}{\widetilde{v}_{0,st}}\, {\bm w}_{1}
\right],
\end{eqnarray}
where $\bar{L}_{B}$ is the adjoint of $L_{B}$ and it is given by
Eq. (\ref{4.16}). The above equation is the same as Eq.\
(\ref{4.14}).

\section{Green-Kubo expression for self-diffusion from the Chapman-Enskog results}
\label{ap2} In ref. \cite{BRCyG00}, the self-diffusion coefficient
of a dilute granular fluid was obtained from the Boltzmann-Lorentz
by the Chapmann-Enskog method. Equations (24) and (26) in the
above-mentioned reference are
\begin{equation}
\label{b1} D=-\frac{1}{d} \int d{\bm v}\, {\bm v} \cdot {\bm
B}({\bm v}),
\end{equation}
where ${\bm B}({\bm v})$ is the solution of the integral equation
\begin{equation}
\label{b2} \left(
K_{BL}+\zeta_{HCS}T_{HCS}\frac{\partial}{\partial T_{HCS}} \right)
{\bm B}({\bm v})=\frac{1}{n_{H}} {\bm v} f_{HCS}({\bm v}),
\end{equation}
$K_{BL}$ being the inelastic Boltzmann-Lorentz collision operator
\begin{equation}
\label{b3} K_{BL}({\bm v})= \sigma^{d-1} \int d{\bm v}_{1} \int d
\widehat{\bm \sigma}\, \Theta ( {\bm g} \cdot \widehat{\bm
\sigma}) {\bm g} \cdot \widehat{\bm \sigma} \left[ \alpha^{-2}
b^{-1}_{\bm \sigma}({\bm v},{\bm v}_{1})-1 \right] f_{HCS}({\bm
v}_{1},t).
\end{equation}
We introduce dimensionless time and velocity scales by
\begin{equation}
\label{b4} s=\int_{0}^{t} dt^{\prime}\,
\frac{v_{0}(t^{\prime})}{\ell}, \quad {\bm c}= \frac{\bm
v}{v_{0}(t)}\, .
\end{equation}
In terms of them, Eq.\, (\ref{b2}) becomes
\begin{equation}
\label{b5} \left( {K}_{BL,c}-\frac{\zeta_{0}}{2}
\frac{\partial}{\partial {\bm c}} \cdot {\bm c} \right)
\widetilde{\bm B} ({\bm c})={\bm c} \chi_{HCS}({\bm c}),
\end{equation}
with
\begin{equation}
\label{b6} \widetilde{\bm B}({\bm c}) =\frac{v_{0}^{d}(t)}{\ell}\,
{\bm B}({\bm v})
\end{equation}
and
\begin{equation}
\label{b7} K_{BL,c}({\bm c})=\frac{\ell}{v_{0}(t)} K_{BL}({\bm
v}).
\end{equation}

The formal solution of Eq. (\ref{b5}) can be written as
\begin{equation}
\label{b8} \widetilde{\bm B} ({\bm c})= - \int_{0}^{\infty} ds\,
e^{s \Lambda_{BL,c}({\bm c})} {\bm c} \chi_{HCS}({\bm c}),
\end{equation}
\begin{equation}
\label{b9} \Lambda_{BL,c}({\bm c})=K_{BL,c}({\bm c})
-\frac{\zeta_{0}}{2} \frac{\partial}{\partial {\bm c}} \cdot {\bm
c}.
\end{equation}
Substitution of Eq.\ (\ref{b8}) into Eq.\ (\ref{b1}) gives
\begin{eqnarray}
\label{b10} D & = & \frac{v_{0}(t) \ell}{d} \int_{0}^{\infty} ds
\int d{\bm c}\, {\bm c} \cdot e^{s \Lambda_{BL,c}({\bm c})} \left[
{\bm c} \chi_{HCS}({\bm c}) \right] \nonumber \\
&=& \frac{v_{0}(t) \ell}{d} \int_{0}^{\infty} ds \int d{\bm c}\,
\left[ e^{s \bar{\Lambda}_{BL,c}({\bm c})}  {\bm c} \right]
 \cdot {\bm c}\chi_{HCS}({\bm c}),
\end{eqnarray}
where $\bar{\Lambda}_{BL,c} ({\bm c})$ is the adjoint of
$\Lambda_{BL,c}({\bm c})$,
\begin{equation}
\label{b11} \bar{\Lambda}_{BL,c}({\bm c}_{1})= \int d {\bm
c}_{2}\, \chi_{HCS} ({\bm c}_{2}) \int d\widehat{\bm \sigma}\,
\Theta ({\bm c}_{12}\cdot \widehat{\bm \sigma}) {\bm c}_{12}\cdot
\widehat{\bm \sigma}[b_{\bm \sigma}({\bm c}_{1},{\bm c}_{2})-1] +
\frac{\zeta_{0}}{2} {\bm c}_{1} \cdot \frac{\partial}{\partial
{\bm c}_{1}}\, .
\end{equation}

\end{document}